\documentclass[twocolumn]{aastex631}

\usepackage{comment}
\usepackage{graphicx}	% Including figure files

\usepackage{amsmath}	% Advanced maths commands
\usepackage{amssymb}	% Extra maths symbols
\usepackage{CJK}
\usepackage{hyperref}
\usepackage{color}
\shorttitle{Population of Massive AGN Stars}
\shortauthors{Chen \& Lin}
\begin{document}

\title{The Population of Massive Stars in Active Galactic Nuclei Disks}

%Dynamical Evolution, Effective Viscosity and Luminosity Heating}

%\correspondingauthor{August Muench}
%\email{greg.schwarz@aas.org, gus.muench@aas.org}

\author[0000-0003-3792-2888]{Yi-Xian Chen}
\affiliation{Department of Astrophysical Sciences, Princeton University, USA}

\author{Douglas N. C. Lin}
\affiliation{Department of Astronomy and Astrophysics, University of California, 
Santa Cruz, USA}
\affiliation{Institute for Advanced Studies, Tsinghua University, China}



\begin{abstract}
Gravitational instability in the outskirts of Active Galactic Nuclei (AGN) disks lead to disk fragmentation and formation of 
%$\sim 300M_\odot$ 
super-massive (several $10^2 M_\odot$) stars with potentially long lifetimes. 
Alternatively, 
stars can be captured ex-situ and grow from gas accretion in the AGN disk. 
However,
the number density distribution throughout the disk is limited by thermal feedback as their luminosities provide the dominant heating source. 
We derive equilibrium stellar surface density profiles under two limiting contexts:
in the case where the stellar lifetimes are prolonged due to recycling of hydrogen-rich disk gas, 
only the fraction of gas converted into heat is removed from the disk accretion flow. 
Alternatively, if stellar composition recycling is inefficient and stars can evolve off the main sequence, 
the disk accretion rate is quenched towards smaller radii resembling a classical star-burst disk, 
albeit the effective removal rate depends not only on the stellar lifetime, but also the mass of stellar remnants. 
For AGNs with central Supermassive Black Hole (SMBH) masses of $\sim 10^{6} - 10^{8} M_\odot$ accreting at {$\sim 0.1$ Eddington efficiency, 
we estimate a total number of $10^3-10^5$ coexisting massive stars and the rate of stellar mergers to be $ 10^{-3} - 1 $ per year}.
We motivate the detailed study of interaction between a swarm of massive stars through hydro and N-body simulations to provide better prescriptions of dynamical processes in AGN disks, 
and to constrain more accurate estimates of the stellar population.
\end{abstract}



\section{Introduction} \label{sec:intro}

Supermassive Black Holes (SMBH) 
are known to host accretion disks that provide power to active galactic nuclei (AGN) in the centers of massive galaxies \citep{Lynden-Bell1969}. 
Strong radiative cooling in the outer regions of these disks lead to fragmentation via gravitational instability \citep{Goodman2003,GoodmanTan2004}, 
especially in a radiation pressure dominated environment \citep{Jiang11,Chen2023}. 
Subsequently, 
the fragments can grow from collision and gas accretion to become massive stars. 
Alternatively, 
stars in the nuclear cluster can be captured onto the disk and continue to grow by gas accretion \citep{Artymowicz1993,MacLeod2020,Wang2023}. 

The onset of in-situ fragmentation, 
or star formation, 
is associated with the fact 
that heating by MRI and gravito-turbulence are 
insufficient to maintain energy balance against radiative cooling. 
Assuming star formation provides the necessary auxiliary power to maintain energy equilibrium at marginal gravitational instability (Toomre parameter $Q\sim 1$), 
\citet{TQM} constructed a star-burst disk model in which 
the accretion rate $\dot{M}_{\rm d}$ should decrease towards the SMBH as mass flux is continuously dumped into stars.

However, considering that stars embedded in AGN disks can form from massive self-gravitating clumps or grow up to a few hundred $M_\odot$ from accretion of external gas, 
their evolution and fates can be significantly different from field stars. 
In particular, 
recent works have pointed out that since the accretion disk evolves on a much longer Salpeter timescale ($\sim 10^8$ years), 
embedded massive stars may replenish their hydrogen reservoir by recycling fresh gas from the disk within a few parsecs of the disk \citep{Cantiello2021,Jermyn2022}. 
Consequently, they undergo little chemical/nuclear evolution and may stay on the 
hydrogen-burning 
main sequence for a timescale that is comparable to AGN lifetime $\tau_{\rm AGN}$.
These ``immortal stars" (IMS) become efficient nuclear power plants that burns whatever 
amount of hydrogen they accrete from the ambient gas into heat and return all helium 
byproducts back into the disk. 
In fact, 
\citet{Jermyn2022} focused their analysis of massive stars to the inner $\sim 0.03$ pc (since that is where disk-capture mechanism is most efficient) 
and already infer an average count of  $\sim 10^2-10^4$ coexisting immortal stars within an AGN disk, 
while it should be further realized that star formation is in fact capable of producing massive stars over a wider range of distances \citep{Jiang11,Chen2023}. 
In the IMS scenario, after the formation of the first generation of stars, 
only a small fraction of gas accreted on to these stars is converted into heat to balance radiative cooling, 
while the rest is recycled out back into the disk, leading to a negligible reduction in disk accretion rate.

Nevertheless, the IMS track may not be the whole picture. 
Supporting
evidences for ongoing stellar and chemical evolution in AGN disks
are provided by the observed red-shift independent super solar
metallicity of AGNs' broad line regions and their enhanced heavy-element 
abundance, relative to the narrow emission line regions \citep{Huang2023}. 
These properties are inconsistent with the implication of ``immortal
stars'' powered by CNO burning on the main sequence without
any net heavy element production.  However, if the accreted gas is buffered 
by a radiative envelope and is not well mixed with the nuclear burning zone 
deep in the stellar interior, the accumulation of He ashes from CNO burning 
would lead to the onset of efficient He burning, the transition to post main 
sequence evolution, and eventually collapse into black holes or supernova 
explosion with prolific heavy-element yields \citep{alidib2023}. 
In this ongoing ``Stellar Evolution and Pollution in AGN Disk'' (SEPAD) 
track,  massive stars still accrete and radiate just like the IMS model, 
but the overall efficiency of stars in extracting gas density from the disk to convert into 
luminosity is different due to their limited lifespan, 
and in turn the disk accretion rate 
is regulated in a different manner.

The coexistence and concentration of a large population of massive 
stars, along either the IMS or SEPAD tracks,  raises the probability 
of stellar mergers.  Stellar coalescence events may continuously increase individual stars' mass and 
luminosity. If this process proceeds rapidly, an energy equilibrium would 
not be sustainable at all times. 

In this paper, 
we study the population of massive stars in an AGN based on consistent disk models at thermal equilibrium, 
in order to provide constraints on their number distribution and merger rates. 
We formulate the main equations for AGN disk models and stellar heating in \S 
\ref{sec:formulation}. 
We quantify key differences between 
IMS and SEPAD tracks and their influences to the disk 
structure. 
In \S \ref{sec:dyn}, we estimate 
the dynamical merger timescales of stellar coalescence as functions of disk and star properties. In \S \ref{sec:diskmodels}, 
we solve specific disk models in the IMS and SEPAD limits as well as the merger timescales associated with their stellar density profiles. 
In \S \ref{sec:conclusion}, 
we summarize our findings and discuss future prospects 
for subsequent numerical work that could refine our model of massive star and their impact on the AGN disk.

\section{Linking Stellar population with Disk Structure}
\label{sec:formulation}

\subsection{General equations}

We illustrate our problem based on a 
parameterized AGN disk model, 
similar to what is applied in \citet{MacLeod2020,Davies2020}. 
A SMBH of mass $ M_{\bullet}$ is fed by a mass flux of

\begin{equation}
    \dot{M}_\bullet = \lambda_\bullet \dot{M}_{\rm Edd} = 0.22 \lambda_\bullet  
    m_8 \epsilon_\bullet ^{-1} M_\odot \text{yr}^{-1}
\label{eq:mdotbul}
\end{equation}
where $m_8 \equiv M_{\bullet}/10^8 M_\odot$, $\dot{M}_{\rm Edd} \equiv L_{\rm Edd}/\epsilon_\bullet c^2$ 
is the Eddington mass accretion rate. 
$\lambda_\bullet \equiv L/L_{\rm Edd}$ is the normalized 
luminosity and $\epsilon_\bullet$ parameterizes the accretion efficiency. 

SMBHs are fed by their surrounding disks at a rate ${\dot M}_{\rm d} $ 
determined by the efficiency of angular momentum transfer 
\citep{lyndenbell1974, pringle1981}. 
With the conventional 
$\alpha_\nu$ prescription \citep{1973A&A....24..337S} for viscosity 
($\nu = \alpha_\nu c_{\rm s}^2/\Omega $ where $\alpha_\nu$ is a 
factor representing angular momentum transport efficiency, 
$c_{\rm s}$ is the sound speed, 
$\Omega=\sqrt{G M_\bullet /R^3}$, 
$P= 2 \pi/\Omega$ are the Keplerian angular 
velocity and orbital period, 
at a distance of $R$, 
respectively. 
The disk accretion rate can be written as

\begin{equation}
{\dot M}_{\rm d}  = 3 {\sqrt 2} (\alpha_\nu h^3/Q) M_{\bullet} 
\Omega 
\label{eq:mdotdisk}
\end{equation}
where $Q \equiv c_s \Omega / \pi G \Sigma$ is the gravitational 
stability parameter\citep{safronov1960, toomre1964}. The aspect ratio is
$h = H/R$ where $H$ is the scale height. The radial velocity is given by

\begin{equation}
V_{\rm R} = 3\alpha_\nu h^2 \Omega R/2.
\label{eq:vrviscous}
\end{equation}

If accretion is driven by magneto-rotational
instability (MRI) with viscous $\alpha \lesssim 10^{-2}$ \citep{balbus1991, 
balbus1998, bai2011, deng2019} and the radiative cooling is balanced by accretion heating alone,
gravitational instability (GI) would occur (with $Q \lesssim 1$) in 
region with $R  \gtrsim R_{ Q = 1} \simeq 10^{-2}$ pc
around SMBHs with $m_8 \sim 0.01-1$ \citep{paczynski1978}. 
GI leads to gravito-turbulence with $\alpha_\nu \sim 0.01-1$ 
\citep{lin1987, lin1988, kratter2008, zhu2010a, zhu2010b, martin2011, 
rafikov2015, deng2020} and prolific rates of {\it in situ} star formation 
\citep{Goodman2003, GoodmanTan2004}.

In star-forming disk regions of interest, viscous heating from turbulence 
is much weaker than radiative cooling, and an energy balance needs to be 
maintained by radiation from either a permanent population of IMS stars or the persistent formation of SEPAD stars.
The latter mode may lead to a strongly (radially) variable $\dot{M}_{\rm d}$ (see \S \ref{sec:dyn}). 
Nevertheless, 
we still expect the disk to self-regulate and maintain $Q\sim 1$ everywhere
due to the requirement for a continuous replenishment of heat sources. 
The viscous stress 
becomes decoupled from the energy equation and is simply a parametrization of the inflow velocity $V_{\rm R}$ from 
a maximum level of turbulence that the disk is able to support. 

Assuming $Q=1$ and ${\dot M}_{\rm d} \simeq 2 \pi \Sigma V_R R$ at all $R$, 
we determine the aspect ratio and surface density:

\begin{equation}
    h = \left (\dfrac{2}{3\alpha_\nu}\dfrac{\dot{M_{\rm d}}}{{M_\bullet}\Omega}\right)^{1/3} 
\label{eq:diskh}
\end{equation}

\begin{equation}
\Sigma = \left( {2 {\dot M}_{\rm d} \over 3\alpha_\nu \Omega} \right)^{1/3} \dfrac{{M_\bullet}^{2/3} }{\pi R^2 }
\label{eq:disksig}
\end{equation}
In outer regions, the sound speed $c_{\rm s} = h V_{\rm K}$ becomes flat for local instability with a given $\alpha_\nu$ and ${\dot M}_{\rm d} $ \citep{SirkoGoodman2003}, 
where $V_{\rm K}=\Omega R$ is the Keplerian velocity.

For a vertically self-similar disk model, the mid-plane density 

\begin{equation}
\rho_{\rm c} \approx \Sigma/2 h R, 
\end{equation}
while the midplane characteristic temperature $T_{\rm c}$ satisfies the Eddington quartic when optically thick

\begin{equation}
      \rho_{\rm c} c_{\rm s}^2 = P_{\rm gas} + P_{\rm rad}  = \rho_{\rm c} \dfrac{RT_{\rm c}}{\mu} +\dfrac13 a T_{\rm c}^4
\end{equation}

As we show below, 
the outer regions of such disks are self-consistently radiation dominated, 
since radiation to gas pressure fraction is an increasing function of $r$ 
(the relation is \textit{reversed} if turbulence provides heating in a constant $\alpha_\nu$, gravitationally stable standard disk \citep{1973A&A....24..337S}, 
which is beyond the scope of our paper).

In dense regions where the disk is optically thick, 
we can express the radiative cooling rate from two sides of the disk $Q^-=2 \sigma T_{\rm eff}^4$ 
(to be differentiated with the Toomre $Q$ parameter) as a function of 
these variables \citep{Hubeny1990,Jiang11,Chen2023}:

\begin{equation}
     Q^{-} = \dfrac{32\sigma T_{\rm c}^4}{3\kappa \Sigma}
     \label{eqn:Q-}
\end{equation}
where $\kappa$ is a constant opacity set to 
order unity in cgs unit. 
{This simple prescription 
approximates
the grain opacity 
within $10^2-10^3$ K 
and the electron opacity above $10^4$K \citep{BellLin1994}. 
We discuss 
the possible effect of an opacity window close to the grain sublimation temperature in \S \ref{sec:mass_budget}}. 
Interpolation between optically thin and optically thick expressions yield a more general expression \citep{SirkoGoodman2003}

\begin{equation}
    Q^{-} = \dfrac{32\sigma T_{\rm c}^4}{3\kappa \Sigma + 8 + 8/\kappa \Sigma}.
     \label{eqn:Q-full}
\end{equation}
The radiation pressure is accordingly generalized as 

\begin{equation}
    P_{\rm rad} = \dfrac{a T_{\rm c}^4}{3\kappa \Sigma + 8 + 8/\kappa \Sigma} {\kappa\Sigma}
\end{equation}

\subsection{Embedded stars as auxiliary heating sources}

The outer regions of marginally (un)stable ($Q=1$) disks are prone to fragmentation and star formation. 
To provide thermal feedback that re-establishes energy equilibrium, stars 
convert some portion of the gas mass into heating, through CNO during 
the main sequence or triple-$\alpha$ reactions during the post-main sequence 
evolution. 
In contrast to star forming regions in galaxies, the emerging 
stars in AGN disks continue to accrete gas on both the IMS and SEPAD tracks. 
As these embedded stars gain mass, they also release intense winds. 
In most regions of AGN disks, 
their growth is suppressed by an accretion-wind balance when they attain equilibrium masses $M_\star (\simeq 10^3 M_\odot)$ with luminosities $L_\star (\simeq 10^7 L_\odot)$ comparable to their own Eddington limit $L_{\rm Edd, \star}$ \citep{Cantiello2021, Jermyn2022}. 
Assuming one characteristic stellar mass $M_\star$ and luminosity $L_\star$, 
the heating term in an equilibrium state is fundamentally given by 

\begin{equation}
     Q^+_\star = s_{\star} L_{\star}
    \label{eqn:initial_SF}
\end{equation}
where $s_{\star}$ is the surface number density of stars. 
An energy equilibrium is reached with 

\begin{equation}
Q^+ _\star \simeq Q^-, 
\label{eq:energyequi}
\end{equation} 

However, depending on the evolution tracks of stars and details of dynamical interactions, 
a fixed amount of heating rate corresponds 
to different reduction rates of gas supply from the disk accretion flow.
The disk surface density evolves with a sink term

\begin{equation}
    {\partial \Sigma \over \partial t} + {1 \over R} {\partial \over \partial R} \Sigma V_{\rm R} R
    = -\dot{\Sigma}_{\rm net}
    \label{eq:sigmaevolve}
\end{equation}
The net removal rate $\dot{\Sigma}_{\rm net}$ is linked to 
$ Q^+_\star$ through several terms,

\begin{equation}
    \dot{\Sigma}_{\rm net} = 
    \dot{\Sigma}_{\rm energy} + \dot{\Sigma}_{\rm rem}   + \dot{\Sigma}_{\rm BH}  
    \label{eq:sigmadiffusion}
\end{equation}

The conversion of rest-mass energy into radiation through nuclear burning
is given by $\dot{\Sigma}_{\rm energy} = s_{\star} L_{\star}/c^2$.
In an accretion-wind equilibrium 
\citep{Cantiello2021}, 
the disk-star mass exchange on the 
IMS and SEPAD tracks do not contribute to ${\dot \Sigma}_{\rm net}$.
But, the formation of black-hole remnants at the end of stars' 
lifespan $\tau_\star$ effectively permanently removes each remnants mass ($M_{\rm rem}$)
from the disk-gas reservoir at a rate 

\begin{equation}    
\dot{\Sigma}_{\rm rem} = s_{\star} M_{\rm rem}/\tau_{\star}.
\label{eq:sigremdot}
\end{equation} 

\begin{figure}[htbp]
\centering
\includegraphics[width=0.42\textwidth,clip=true]{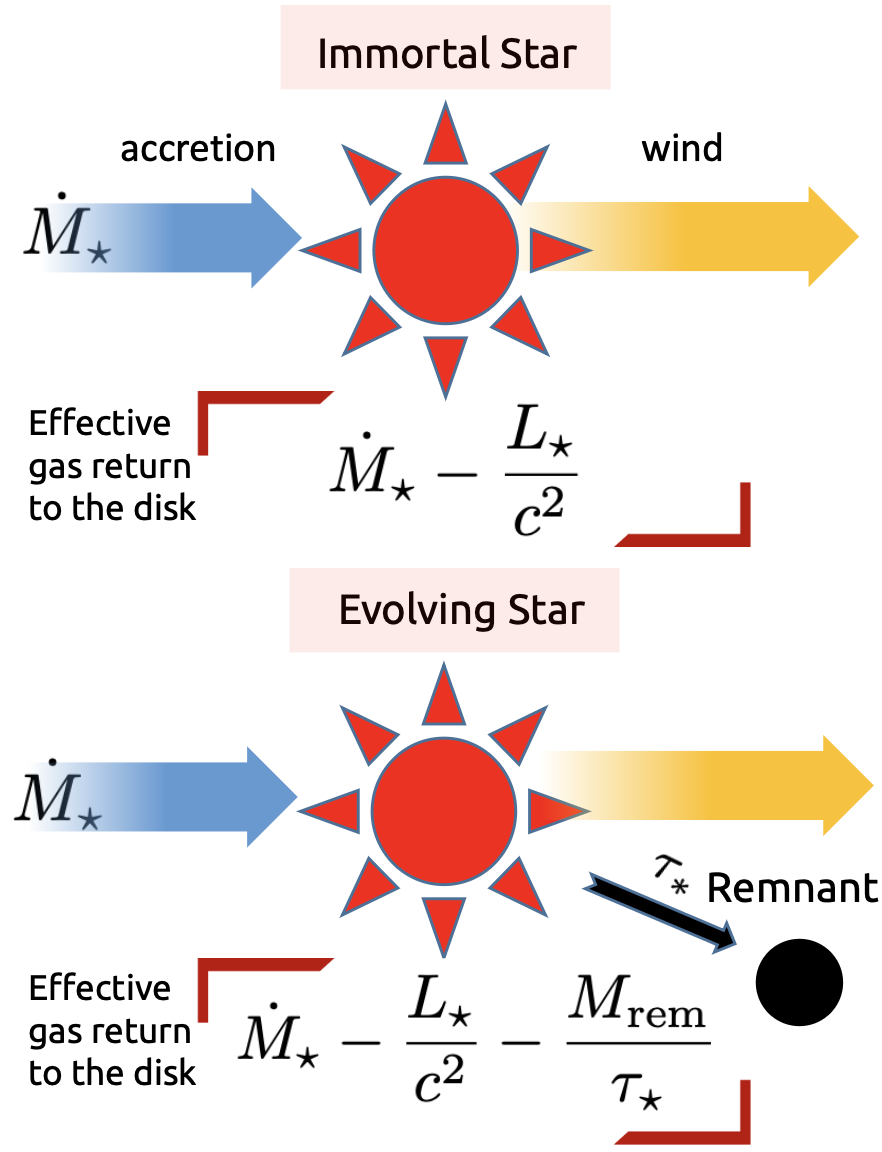}
\caption{{Differences between the IMS and the SEPAD stellar evolution models. 
For immortal stars, 
all gas deposited into stars is eventually recycled out, save for the fraction consumed to generate radiative energy. 
For stars that evolve off the main sequence within an evolution timescale of $\tau_\star$, 
there is an additional removal rate of $M_{\rm rem}/\tau_\star$.}}
\label{fig:schematics}
\end{figure}

{The effective channels of gas return to the disk 
for two stellar evolution scenarios 
are shown in the schematics of Figure \ref{fig:schematics}.}

At the end of the SEPAD track, 
the magnitude of $M_{\rm rem}$ may be a few $M_\odot (\ll  M_\star)$ and the remaining mass $M_\star - M_{\rm rem}$ is returned to the disk flow,
while a new generation of stars with mass $M_\star$ forms to fill the energy deficit on the dynamical timescale $\sim \Omega^{-1} \ll \tau_\star$, 
maintaining an equilibrium stellar surface density. 
As black-hole remnants accumulate in number and accrete mass at the Eddington-limited
rate, they deplete mass from the gas at a rate per unit area
\begin{equation}   
{\dot \Sigma}_{\rm BH} (t) = \Sigma_{\rm BH} (t) /\tau_{\rm Sal}
\end{equation}
where $\tau_{\rm Sal}$ is the Salpeter timescale and $\Sigma_{\rm BH} (t)$ 
is the surface mass density of remnant black-holes. Since $\tau_{\rm Sal}$ 
is generally much longer
than other timescales, 
contributions
from ${\dot \Sigma}_{\rm BH}$ ramps slowly with $t$ and becomes significant
towards the advanced stages (at $\sim \tau_{\rm AGN}$) of AGN evolution. 
On the other hand, in the IMS limit, neither  $\dot{\Sigma}_{\rm rem}$ nor $\dot{\Sigma}_{\rm BH}$ exists.

\subsection{Channels of gas depletion}

Neglecting the $\dot{\Sigma}_{\rm BH}$ term first in a time-independent calculation, 
Equation  (\ref{eq:sigmadiffusion}) reduces to \begin{equation}
    \dot{\Sigma}_{\rm net} = \dfrac{s_* L_*}{c^2}\left(1 + \dfrac{\tau_0}
    {\epsilon_\star M_{\star}}\dfrac{M_{\rm rem}}{\tau_*}\right)
    \label{eq:sigmadotnet}
\end{equation}
where $\epsilon_\star \sim 10^{-3}$ is the efficiency factor defined from
the stars' characteristic evolution timescale if evolved in isolation
$\tau_0 \equiv \epsilon_\star M_\star c^2/L_\star$,   which is 
$\ll \tau_\star$ for the IMS stars 
(due to stellar metabolism) 
and $\tau_0 \sim \tau_\star$ for the SEPAD stars. 
To first order, 
the influence of feedback on the accretion flow from different stellar evolution models can all be expressed in terms of one gas-recycle efficiency factor 

\begin{equation}
\epsilon_{\rm recycle} 
:= s_\star L_\star / \dot{\Sigma}_{\rm net} {c^2} = \left(1 + \dfrac{\tau_0}
    {\epsilon_\star M_{\star}}\dfrac{M_{\rm rem}}{\tau_*}\right)^{-1},
\label{eqn:epsilon}
\end{equation}
the net gas
removal rate is small when gas is efficiently recycled and $\epsilon_{\rm recyle} = 1$ as in the IMS case. 
For stars on the SEPAD track, 
$\epsilon_{\rm recycle}\sim \epsilon_\star M_\star/M_{\rm rem}$. 
Note when $M_\star/M_{\rm rem}\sim 1$ 
as the average property of field stars, 
$\epsilon_{\rm recycle}\sim \epsilon_\star$ reduces to the \citet{TQM} model, 
but such a case might not be appropriate 
for the SEPAD stars
of typically more than a hundred solar masses (\S\ref{sec:intro}).

With these prescriptions in a thermal equilibrium ($s_\star L_\star= Q^+ _\star = Q^-$), 
Eqn \ref{eq:sigmaevolve} reduces to

\begin{equation}
    \dot{M}_{\rm d} (R_{\rm out}) - \dot{M}_{\rm d} =2\pi  \int_R^{R_{\rm out}} \dfrac{Q^-}{\epsilon_{\rm recycle} c^2} R' {\rm d} R'.
    \label{eqn:variablemdot}
\end{equation}

Besides these effects, 
coexisting stars also undergo close encounters, binary capture, 
orbital in-spiral and coalescence, 
which may bring complication in the effective recycle frequency. 
For massive stars reaching equilibrium mass, their coalescence is followed by shedding of one star's worth of mass to the disk during a short shedding timescale \citep[][also see \S \ref{sec:shedding}]{Jermyn2022}. 
If this shedding timescale is much shorter than the typical merger timescale $\tau_{\rm shed} \ll \tau_{\rm merge}$, 
the stellar properties maintain their equilibrium values. 
For the IMS, 
this means no net change in the expression for $\epsilon_{\rm recycle}$ since an immortal star remains immortal after merger. 
For SEPAD track, 
the mixing of composition between young and old stellar population may introduces subtle changes in the effective $\tau_\star$ which affects $\epsilon_{\rm recycle}$. 

On the other hand, 
if merger is very frequent and $\tau_{\rm shed} \gtrsim  \tau_{\rm merge}$, 
AGN stars would undergo runaway growth from sequential mergers and result in supermassive star that
eventually explode as pair instability supernovae or collapse into massive black holes, 
but we show below that this scenario is rather unlikely since merger efficiency is limited by the timescale of inspiral after stars are captured into binaries.
In the next section, 
we discuss how to estimate the merger timescale based on given disk and star properties.

\section{Estimation of Merger Timescale}
\label{sec:dyn}

%In this section, we calculate the effective merger timescale $\tau_{\rm merger}$, which provides an estimate of stellar merger event rates in preservative IMS and conservative SEPAD models. In the lossy IMS and SEPAD models, this timescale is also coupled with disk evolution through $\epsilon_{\rm recycle}$.

\subsection{Stellar coalescence}
\label{sec:capture}

\subsubsection{Binary capture timescale}
\label{sec:capture_timescale}

As a large number of massive stars co-exist in the disk, they encounter each other through dynamical interactions. 
High stellar surface number density may result in a short merger timescale, 
$\tau_{\rm merger}$ within which most stars can go through one typical 
merger process. 
{The stellar capture process is analogous 
to planetesimal coagulation  \citep[e.g.][]{idalin2004}. 
In particular, without gas damping, 
the physical collisional
frequency is based on the radius of stars $R_{\star}$ enhanced by gravitational focusing \citep{binney1987, Palmer1993}}

\begin{equation}
    \Gamma_{\rm col} \simeq \pi R_{\star}^2 (1 + \Theta_\star) 
    {s_\star \over H_\star} \Delta V 
    \sim \pi (1+\Theta_\star) R_\star^2  \Omega,
    \label{eqn:gammacol_gasless}
\end{equation}
with a corresponding timescale of

\begin{equation} 
\tau_{\rm col} 
= {1 \over \Gamma_{\rm col}}
%\sim {1 \over \pi (1+\Theta_\star) R_\star^2  \Omega}
\label{eq:taucol}
\end{equation}
{where $H_\star = \Delta V/\Omega$ is the stars' dynamical scale height, 
$\Theta_\star =
G M_\star/ R_{\star} \Delta V^2$ is the stellar \citet{safronov1972} 
number, $\Delta V$ is star's dispersion velocity amplitude (\S\ref{sec:vdisp}).}

{For gas-poor planetary
rings \citep{goldreich1978, stewart1984} and planetesimals 
\citep{Palmer1993}, 
the effective capture impact parameter $\sim (1+ \Theta)^{0.5} R_\star$ 
is usually much smaller than the Hill radius of these objects. 
This is because if planetesimals start from a cold configuration, 
their dispersion (non-circular) 
velocity $\Delta V$ 
grows while their $\Theta_\star$ 
decreases 
on a disk-star-relaxation timescale \citep{Palmer1993, Aarseth1993} }

\begin{equation}
    \tau_{\rm V +} \simeq { \tau_{\rm col} \over (1+ \Theta_\star)}
\label{eq:tauv+}
\end{equation}

{On the collisional timescale ($\tau_{\rm col}$), 
stars dissipate
their kinetic energy at a rate $d \Delta V^2 / dt \sim \Delta V^2/\tau_{\rm col}$,
i.e. the collisional damping timescale $\tau_{\rm damp} \sim \tau_{\rm col}$. 
Therefore, 
a dynamical equilibrium is established with 
$\tau_{\rm col} \sim \tau_{\rm V +}$ 
or $\Theta \sim 1$. }

{In gas-rich disks, 
embedded stars are 
surrounded and fed by mini-disks.  
Simulations have shown that during their 
encounters, 
disk-embedded binaries can become bound after their separation 
reach of order $R_{\rm H} (\equiv (M_\star/3 M_\bullet)^{1/3} 
R \gg R_\star$ ) \citep{Wang2018,
LiLaiRodet2022,Jiaruli2023}, 
which means their capture-impact parameter is effectively
$ \sim R_{\rm H}$. 
This is because the existence of gas torques tend to damp velocity dispersion 
towards a much lower value corresponding to $\Theta \gg 1$, 
such that the effective impact parameter
grows towards $R_{\rm H}$ for increasing gas density. 
However, it 
must be capped at $R_{\rm H}$ when the velocity amplitude is consistently reduced 
down to the Hill velocity $V_{\rm H} = \Omega R_{\rm H}$. 
To connect the gas-poor and gas-rich regimes, 
it is useful to introduce }

\begin{equation}
    R_{\rm eff}^2  = \max[({1+\Theta_\star}) R_\star^2  , R_{\rm H}^2 ],
    \label{eqn:Reff}
\end{equation}
{such that the generalized expression for capture rate and timescale becomes}

\begin{equation}
    \Gamma_{\rm cap} = \pi R_{\rm eff}^2 \Omega, \ \ \ \tau_{\rm cap} = 1/ \Gamma_{\rm cap}.
    \label{eqn:final_capturetimescale}
\end{equation}

\subsubsection{The migration timescale}
\label{sec:mig}

Due to their tidal interaction with the AGN disk, isolated 
low-mass stars are affected by type I migration torque \citep{goldreich1980, 
ward1997} with well-established prescriptions for migration 
timescales 

\begin{equation}
    \tau_{\rm mig, I} \simeq {h Q \over 2 \pi f_{\rm I}} {M_\bullet \over M_\star \Omega} 
    \label{eq:taumig1}
\end{equation}
with $f_{\rm I} \sim {\mathcal O} (1) $ being a dimensionless efficiency factor 
determined by $\Sigma$ and $T_{\rm c}$ distribution as well as the companion mass ratio defined as $q = M_\star/M_\bullet$,
$\alpha_\nu$ and $h$ \citep{paardekooper2010, paardekooper2011,
Kley2012, Baruteau2013} in PPDs. 

Moreover,  stars are subject
to the stochastic torque of the turbulent eddies (with a
turnover timescale $\sim P = 2 \pi/\Omega$) which engulf
them \citep{nelson2005, Baruteau2010, Baruteau2013}. Provided their 
$R_{\rm H} \lesssim H$ and sufficiently high $\alpha$, 
embedded stars undergo random walk (either inward or outward) 
over a distance $\Delta R \simeq R (P/ \tau_{\rm mig})$ per 
orbital period \citep{Wu2024}, 
with a diffusion coefficient 
\begin{equation}  
\chi_{\rm mig} \simeq {\Delta R^2 \over  P}
\simeq \left({ 4 \pi^2 q \over h Q} \right)^2 
{R^2 \over P}.
\label{eq:chimig}
\end{equation}

\iffalse
\begin{equation}
{\Delta R \over R}
\simeq \left( {\Delta t \over P} \right)^{1/2}
{P \over \tau_{\rm mig}}
\end{equation}
during a time interval $\Delta t$ .
\fi

There is also the possibility of partial gap formation
and transition from type I to II migration \citep{Lin1986}.  
For more general range of  $q$, the migration timescale in a laminar disk can be expressed as \citep{Kanagawa2018}:

\begin{equation}
    \tau_{\rm mig} \simeq {h Q \over 2 \pi q \Omega }  (1+ 0.04K), 
    \ \ \ K = {q^2 \over h^5 \alpha}.
    \label{eq:migkana}
\end{equation}

The companion follows rapid type-I migration at $K \ll 1$, 
while for large companion mass $K \gg 1$ the migration pace is greatly reduced due to the reduction of gas density around the companion vicinity by gap-opening effect the \citep{Lin1986}. 
However, 
this formula no longer applies for very deep gaps and massive companions' migration can be stalled or even reversed \citep{Chen2020, Dempsey2021}. 

Nevertheless, the above discussions apply to the migration of isolated stars. 
The net 
torque and $\tau_{\rm mig}$ are modified by the
interference between the wake of multiple coexisting stars \citep{GoodmanRafikov2001}. 
These processes imply that the population of stars/compact objects in AGNs 
is unlikely to be uniformly guided towards a migration trap \citep{Bellovary2016} for merging over a timescale of $\tau_{\rm mig}$. 
Instead, 
their merger rates are ultimately governed by their velocity dispersion at each radius. 
A more precise prescription to account for the effect of migration is to introduce radial advection terms in $s_\star$ in a time-dependent calculation. 
In this paper, we 
merely provide $\tau_{\rm mig}$ to calculate the velocity dispersion $\Delta V$ because it is closely connected to the gas damping timescale.

\subsubsection{Velocity dispersion}
\label{sec:vdisp}

In the AGN context, 
stars are embedded in their natal disks. 
Related to the migration torque (\S\ref{sec:mig}), 
the stars' mean 
eccentricity is also damped by their tidal interaction with their host stars
on a timescale of \citep{Artymowicz1993,Kley2006,Li2019,Chen2021}

\begin{equation}
    \tau_{\rm e} \sim \tau_{\rm mig} h^2 \sim  {h^3 Q \over 2 \pi q \Omega } (1+ 0.04K).
%    \ \ \ \ {\rm and} \ \ \ \ \tau_{\rm e} \sim h^2 \tau_{\rm mig}
\label{eq:taue}
\end{equation}

When $\tau_e$ is significantly shorter than $\tau_{\rm damp}$, 
the extra effect of gas damping becomes more significant to additionally damp 
$\Delta V$ towards a value $\ll \sqrt{GM_\star/R_\star}$ 
and therefore $\Theta_\star \gg 1$. 
In this regime, we can also write Equation (\ref{eq:tauv+}) as

\begin{equation}
   \tau_{V+} \sim {\Delta V^4 \over G^2 M_\star^2 s_\star \Omega}.
\end{equation}

A dynamic 
equilibrium is established 
with $\tau_{\rm V +} \sim \tau_{\rm e}$. 
The solution to this equation, $\Delta V_{\rm e}$, satisfies

\begin{equation}
    {\Delta V_{\rm e} \over V_{\rm k}} \sim \left[ { s_\star R^2 h^3 Q  q \over 2 \pi } (1+ 0.04K) \right]^{1/4},
\end{equation}

{This term can be either substituted into $R_{\rm eff}$ or 
directly compared with $V_{\rm H}$ to ascertain the regime that Equations (\ref{eqn:final_capturetimescale}) \& (\ref{eqn:Reff}) belong to. 
If $\Delta V_{\rm e} > V_{\rm H}$, 
the effective gas damping is not yet sufficient to raise $R_{\rm eff}$ to $R_{\rm H}$. 
However, 
when $\Delta V_{\rm e} < V_{\rm H}$, 
the capture radius has already reached its upper limit $R_{\rm H}$ and $\Delta V_{\rm e}$ in fact decouples from the capture timescale. }

\subsection{Inspiral of binary stars}
\label{sec:inspiral}

The gas rich capture process with $R_{\rm eff} \gg R_\star$ 
is much more efficient than direct collision due to much larger cross section. 
However, 
the expense is that even after capture into bound binaries, 
the stars still need to go through an extra inspiral timescale $\tau_{\rm ins}$ towards merger, 
so the effective timescale of one entire merger process is $\tau_{\rm merge} \approx \tau_{\rm ins} + \tau_{\rm cap}$.

Due to the technical challenges of demanding numerical resolution over large dynamical range ($R_{\rm H}/R_\star$),
past numerical simulations of embedded binary in disks around SMBH
(or central stars) have been only resolved to a fraction of 
$R_{\rm H}$ \citep{Baruteau2010,  Dempsey2022,LiLai2022,Lia,Lib}. 
They show that the magnitude of $\tau_{\rm ins}$ is highly uncertain, 
depending on the circumbinary disk (CBD) properties as well as the 
eccentricity, inclination of the hierarchical binaries' orbit,
softening parameter of the gravitational potential, 
and the sink-hole radius of the stars or their black hole remnants.

Ignoring uncertainties in the direction of binary movement, 
a statistically-averaged estimate of the magnitude of the inspiral timescale influenced by the CBD gravitational torque satisfies \citep{Baruteau2010,  Dempsey2022,LiLai2022,Lia,Lib}

\begin{equation}
    \tau_{\rm ins} \sim {a_{\rm b} \over {\dot a}_{\rm b}} \sim {M_B \over \dot{M}_B}
\label{eq:tauins}
\end{equation}
where $M_B \sim 2 M_{\rm star}$ is the total mass of the binary and $\dot{M}_B$ is the accretion flow in the CBD towards the center. 
To first order, 
we can assume $\dot{M}_B$ is comparable to the sum of accretion rates onto each of the binary component $2 \dot{M}_\star$. 
Moreover, 
for massive stars reaching wind-accretion equilibrium, the accretion rate is close to the mass loss rate

\begin{equation}
     \dot{M}_\star \sim \dfrac{ R_\star L_\star}{GM_\star}
     \label{eq:mdotwind}
\end{equation}
tapped by a near-Eddington luminosity $L_\star \sim L_{\rm Edd} 
= 3.2 \times 10^4 L_\odot (M_\star/M_\odot)$. In terms of the 
Eddington factor $\lambda_\star= L_\star/L_{\rm Edd}$, this 
translates to an inspiral timescale of 

\begin{equation}
\begin{aligned}
    \tau_{\rm ins} & \approx 
    10^4 \left(\dfrac{M_*}{300M_\odot} \right)^2 
    \dfrac{10 R_\odot}{R_*} \dfrac{10^7 L_\odot}{L_*} {\rm yrs}  \\ & = 
     {10^4 \over \lambda_\star} \left(\dfrac{M_*}{300M_\odot} \right)
    \dfrac{10 R_\odot}{R_*} {\rm yrs} 
    \end{aligned}
\end{equation}
%{\color {red} If we express $L_*$ in terms of $L_{\rm ed}$ (i.e. $\lambda_*=L_*/L_{\rm ed}$), we can get ride of the $M_*$ dependence.}
where $\lambda_\star$ is the star's luminosity Eddington fraction.
Note that this estimate only considers a set of isolated binaries in a laminar disk. 
It's possible for the inspiral to stall at $R_\star < a_{\rm b} < R_{\rm H}$ 
over a long time ($ \gtrsim \tau_{\rm cap}$), making it favorable
for additional-body interaction to either disrupt the 
binary pair \citep{hut1983} or forge hierarchical systems
\citep{portegieszwart2004}. 
Secular interaction, including
the von Zeipel-Lidov-Kozai, 
evection, and eviction resonances
between the binary and SMBH may also lead to eccentricity
excitation to speed up the merger process
\citep{bhaskar2022, bhaskar2023}. When the disk structure is perturbed by turbulence
in the limit $R_{\rm H} \lesssim H$, the spin orientation
of the gas inside their $R_{\rm H}$ may fluctuate on the 
turbulent-eddy's turn over time ($\sim P$)
\citep{chenlin2023} and slow down the inspiral process.

\subsection{Post-merger mass shedding}
\label{sec:shedding}
Following the increases in their mass to ${\tilde M}_\star 
(> M_\star)$, the merger products either become unstable and
collapse, or adjust to a new hydrostatic equilibrium
on a dynamical timescale. 
In the latter case, the 
elevated pressure and temperature in the stellar
nuclei also strongly 
enhances their nuclear burning 
rate to a luminosity ${\tilde L}_\star$ in excess to
the Eddington limit associated with their newly acquired
${\tilde M}_\star$. 
Since $M_\star$ of the progenitors
provides a balance between their 
accretion and mass loss rates, 
the merged byproducts 
with super-Eddington luminosity shed a fraction 
or all of the excess mass (${\tilde M}_\star - M_\star$) 
during the return course to the pre-merger equilibrium 
state. If this mass readjustment does not occur dynamically during the merger episode, we still expect it to proceed on a timescale of \citep[][see their Equation 15]{Jermyn2022}:

\begin{equation}
    \tau_{\rm shed} = \dfrac{GM_*^2}{R_* L_*}
\label{eq:taushed}
\end{equation}
which bears the same expression as $\tau_{\rm ins}$ due to the wind-accretion equilibrium.
The final fate of stars depends on the hierarchy between $\tau_{\rm shed}$ and the merger timescale. 
We discuss this aspect for specific disk profiles in \S \ref{sec:merger_timescales}.

\section{Disk and Stellar Population Profiles}
\label{sec:diskmodels}

\subsection{IMS Track}
\label{sec:ims_disk}

Neglecting the effect of merger on stellar lifetime and remnant masses, 
we provide some quantitative analysis on the distribution of AGN stars for a constant $\epsilon_{\rm recycle}$. 
In particular, 
we can estimate the magnitude 
of $s_\star$ in
IMS or SEPAD models by 
determining the $Q^-$ profile. 
Along the IMS track with $\epsilon_{\rm recycle}=1$, 
hardly any gas density is depleted to generate enough heating for the disk, 
and the disk  $\dot{M}_{\rm d} \approx \dot{M}_\bullet$. 
$Q^-$ can be determined in terms of  
$M_\bullet, \lambda_\bullet, \epsilon_\bullet, \alpha_\nu, \kappa_{\rm cgs}$ (Eqn \ref{eqn:Q-}). 
{In the limit that $Q^-/\epsilon_{\rm recycle} c^2$ is negligible, 
the disk model resembles that of \citet{SirkoGoodman2003}.}
Specifically,  
Equations \ref{eq:diskh} and \ref{eq:disksig} give, 

\begin{equation}
    h = c_{\rm s}/V_{\rm k} = 0.013 m_8^{-1/6} (\alpha_\nu \epsilon_\bullet /\lambda_\bullet )^{-1/3} r_{\rm pc}^{1/2} 
\end{equation}
where $r_{\rm pc} = R/1$pc and

\begin{equation}
    \Sigma = 86  m_8^{5/6} (\alpha_\nu \epsilon_\bullet /\lambda_\bullet)^{-1/3} r_{\rm pc}^{-3/2} {\rm g/cm}^2.
\end{equation}

Note that Equations 
\ref{eq:diskh} and \ref{eq:disksig} are derived under the assumption $Q=1$. 
At small radii $R \lesssim R_{Q=1} \simeq 0.01$pc, 
so a full disk solution should include a transition to conventional models 
(with $Q^- $ balanced by viscous heating  \citep{1973A&A....24..337S,frank2002}). 

In the radiation pressure dominated and optically thick limit, the cooling rate is 

\begin{equation}
Q^{-} = \dfrac{2.1 \times 10^6 m_8^{5/6} }{\kappa_{\rm cgs} (\alpha_\nu \epsilon_\bullet / \lambda_\bullet)^{1/3}r_{\rm pc}^{3/2}} 
{\rm erg/cm^2 s}.
\end{equation}
From Equation \ref{eqn:initial_SF} we see this function 
correspond to a stellar surface density of

\begin{equation}
    s_{*,\rm IMS} =  \dfrac{520 m_8^{5/6}(10^7 L_\odot/L_\star)}{\kappa_{\rm cgs} (\alpha_\nu \epsilon_\bullet/\lambda_\bullet)^{1/3} r_{\rm pc}^{3/2}}  {\rm pc^{-2}}.
    \label{eqn:s_*,IMS}
\end{equation}

\begin{figure*}[htbp]
\centering
\includegraphics[width=1\textwidth,clip=true]{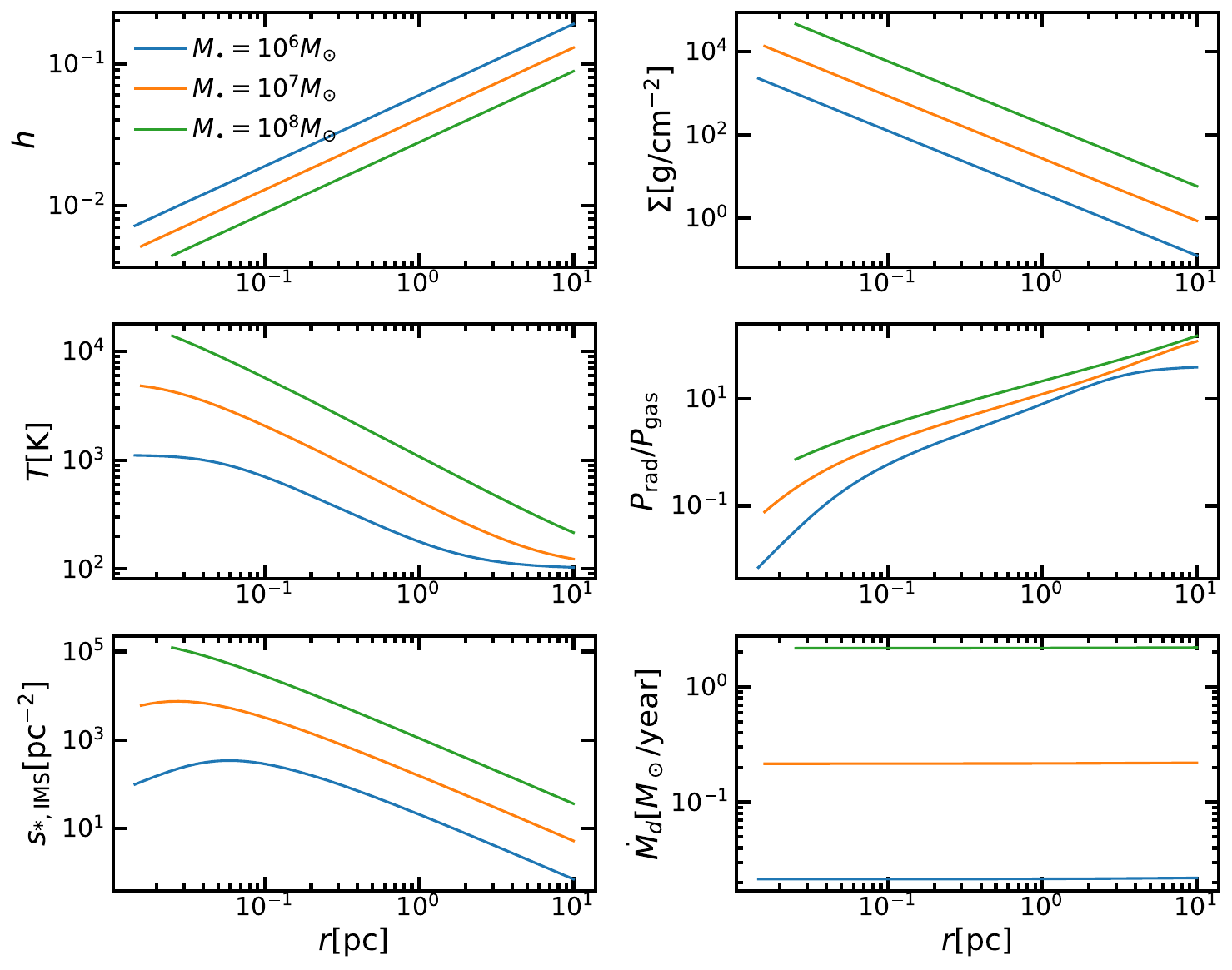}
\caption{The aspect ratio $h$, gas surface density $\Sigma$, 
temperature $T$, 
radiation to gas pressure $P_{\rm rad}/P_{\rm gas}$, equilibrium stellar surface number density $s_{*, \rm IMS} = Q^-/L_\star$, 
as well as the variable disk accretion rate as functions of radius $r_{\rm pc}$ for immortal stars. 
We assume the radiative cooling is balanced by combined luminosity from massive stars with characteristic mass $M_*$ and luminosity $L_*$. 
We assume $\kappa_{\rm cgs} = \alpha_\nu = 1$, and fix $\tilde{\epsilon}_{\rm d}: = \epsilon_{\rm d}/\lambda_{\rm d} = 0.1$ at the outer boundary. 
}
\label{fig:SID_profile}
\end{figure*}

To compare with these scalings, 
we also provide the numerical solutions to the disk equations taking into account 
i) small $\dot{\Sigma}_{\rm net}$ due to $\epsilon_{\rm recycle}=1$; 
ii) both optically thin and thick cooling and iii) both gas and radiation pressure.
The top panels of Figure \ref{fig:SID_profile} show $h$ and $\Sigma$ as functions of $r_{\rm pc}$, 
for different SMBH masses $m_8=0.01,0.1,1$, which 
follow closely the power laws given by Eqns \ref{eq:diskh} and \ref{eq:disksig}. 
We adopt the parameterization 
$\dot{M}_{\rm d}= 0.22 \lambda_{\rm d} m_8 \epsilon_{\rm d}^{-1} M_\odot {\rm yr}^{-1}$ analogous to Eqn \ref{eq:mdotbul} to define a radially varying accretion efficiency 
$\tilde{\epsilon}_{\rm d} = \epsilon_{\rm d}/\lambda_{\rm d}$, 
and {supply an accretion rate 
corresponding to $\tilde{\epsilon}_{\rm d}(R_{\rm out})=0.1$ 
from the outer boundary, 
a value supported by SMBH population models \citep{yu2002,marconi2004,shankar2004,shankar2009} based on optical and x-ray data \citep{soltan1982,fabian1999,elvis2002}}. 
We also adopt $\kappa_{\rm cgs} = \alpha_\nu  = 1$ as our fiducial parameters.
We solve the disk profile from $R_{\rm out}$ inwards and stop at the radius where viscous heating $Q^+_{\nu} = 3\dot{M}_{\rm d}\Omega^2/8\pi$ becomes comparable to $Q^-$, 
since within that radius,
the disk should match 
onto a viscously heated solution with no more star formation.  
The outer boundary is set to be $R_{\rm out} = 10$pc.

In the middle panels of Figure \ref{fig:SID_profile}, 
we show $T_{\rm c}$ and radiation to gas pressure ratio $\Pi (:= P_{\rm rad}/P_{\rm gas})$ 
with molecular weight 
$\mu = 0.6 m_p$. 
At $r_{\rm pc} > r_{\Pi=1} \approx 0.1m_8^{-1/2}$,  
the disk becomes radiation pressure dominated.
The lower panels of 
Figure \ref{fig:SID_profile} show $s_\star$ and disk accretion rates as functions of $r_{\rm pc}$, assuming characteristic $L_\star=10^7
L_\odot$ at their Eddington limits.
At small radius the disk may be gas pressure dominated and $T_{\rm c}$ is a constant, 
meaning $s_\star \propto Q^-\propto \Sigma^{-1}$ increases with radius, so $s_{*,\rm IMS}$ has 
a maximum value at 
$\Pi \approx 1$,

\begin{equation}
    s_{\star, \rm IMS, max} \sim  10^4 \dfrac{m_8^{19/12} (10^7 L_\odot/L_\star)}{\kappa_{\rm cgs} (\alpha_\nu \epsilon_\bullet/\lambda_\bullet)^{1/3}}
     {\rm pc^{-2}}
    \label{eqn:max_density}
\end{equation}

At larger radius where radiation pressure is dominant, $s_{*}$ yields to Eqn \ref{eqn:s_*,IMS} when optically thick. 
Finally, 
the $\dot{M}_d$ profiles in the lower right panel serve as a confirmation that due to large $\epsilon_{\rm recycle}$ 
the reduction in gas flow is negligible and $\dot{M}_d = \dot{M}_\bullet$ is a valid approximation.

\subsection{SEPAD Track}
\label{sec:SEPAD_disk}

In the other limit, 
disk depletion becomes significant due to $\epsilon_{\rm recycle} \ll 1$.  
To estimate typical stellar density profiles under this context, 
instead of approximating with constant $\dot{M}_{\rm d}$, 
one must close 
the Equation set with a variable accretion rate (Eqn \ref{eqn:variablemdot}), 
akin to the procedures of \citet{TQM}. 
For a fixed $\epsilon_{\rm recycle}$
one has the following solution for $\dot{M}_{\rm d}$:

\begin{equation}
    \dfrac{\dot{M}_{\rm d}(R)}{\dot{M}_{\rm d}(R_{\rm out})} = \left[ 1 -  f_{\rm deplete}
    (1 - \sqrt{R/R_{\rm out}}) \right]^{\gamma} 
\end{equation}
where $\gamma$ is an exponent associated with detailed opacity and equation of state scalings. 
The depletion fraction is

\begin{equation}
\begin{aligned}
    f_{\rm deplete} & = f(\gamma) \dfrac{\pi R_{\rm out}^2 Q^-(R_{\rm out})}{\epsilon_{\rm recycle} \dot{M}_{\rm d}(R_{\rm out}) c^2} \\
    & \sim 1.6 
    f(\gamma) \dfrac{m_8^{-1/6} (R_{\rm out}/10{\rm pc})^{1/2}}{\alpha_\nu^{1/3} \tilde{\epsilon}_{\rm d}^{2/3} (\epsilon_{\rm recycle}/0.01)\kappa_{\rm cgs}}
    \end{aligned}
    \label{eqn:f_factor}
\end{equation}
where $f(\gamma)$ is an order-unity factor. 
In the optically thick, 
radiation pressure dominated scenario, 
$\gamma = 3/2$ and 
$f(3/2) = 8/3$. 
Regardless of the exact value of $\gamma$, 
Eqn \ref{eqn:f_factor} represents the fraction of mass that needs to be taken from the disk accretion flow, 
i.e.
subtracted off the accretion rate $\dot{M}_{\rm d}$, 
to support the 
total surface cooling rate $\sim \pi R^2  Q^- $. 
Another equivalent interpretation of $f_{\rm deplete}$ 
is the ratio between star forming timescale ${\Sigma}/\dot{\Sigma}_\star$ and the accretion (disk-clearing) timescale $\pi R^2 \Sigma/\dot{M}_{\rm d}$ \citep{TQM}. 
The $f_{\rm deplete} = 1$ 
condition gives the critical supply,
 ${\dot M}_{\rm crit} = f(\gamma) \pi Q^- (R_{\rm out}) 
R_{\rm out}^2/\epsilon_{\rm recycle} c^2$, 
that is need from the outer boundary for non-negligible fraction of 
$\dot{M}_{\rm d}(R_{\rm out})$ to reach the inner region. 

At $R \ll R_{\rm out}$, 
the accretion rate may relax to a constant 
$ \sim \dot{M}_{\rm d}(R_{\rm out}) (1-f_{\rm deplete})^{\gamma}$ 
for $f_{\rm deplete} < 1$.
Within this radius, 
the stellar density follows

\begin{equation}
    s_{*,\rm SEPAD} \approx   (1-f_{\rm deplete})^{\gamma} f_{L_\star} s_{*,\rm IMS} 
    \label{eqn:s_*_SBD}
\end{equation}

\begin{figure*}[htbp]
\centering
\includegraphics[width=1\textwidth,clip=true]{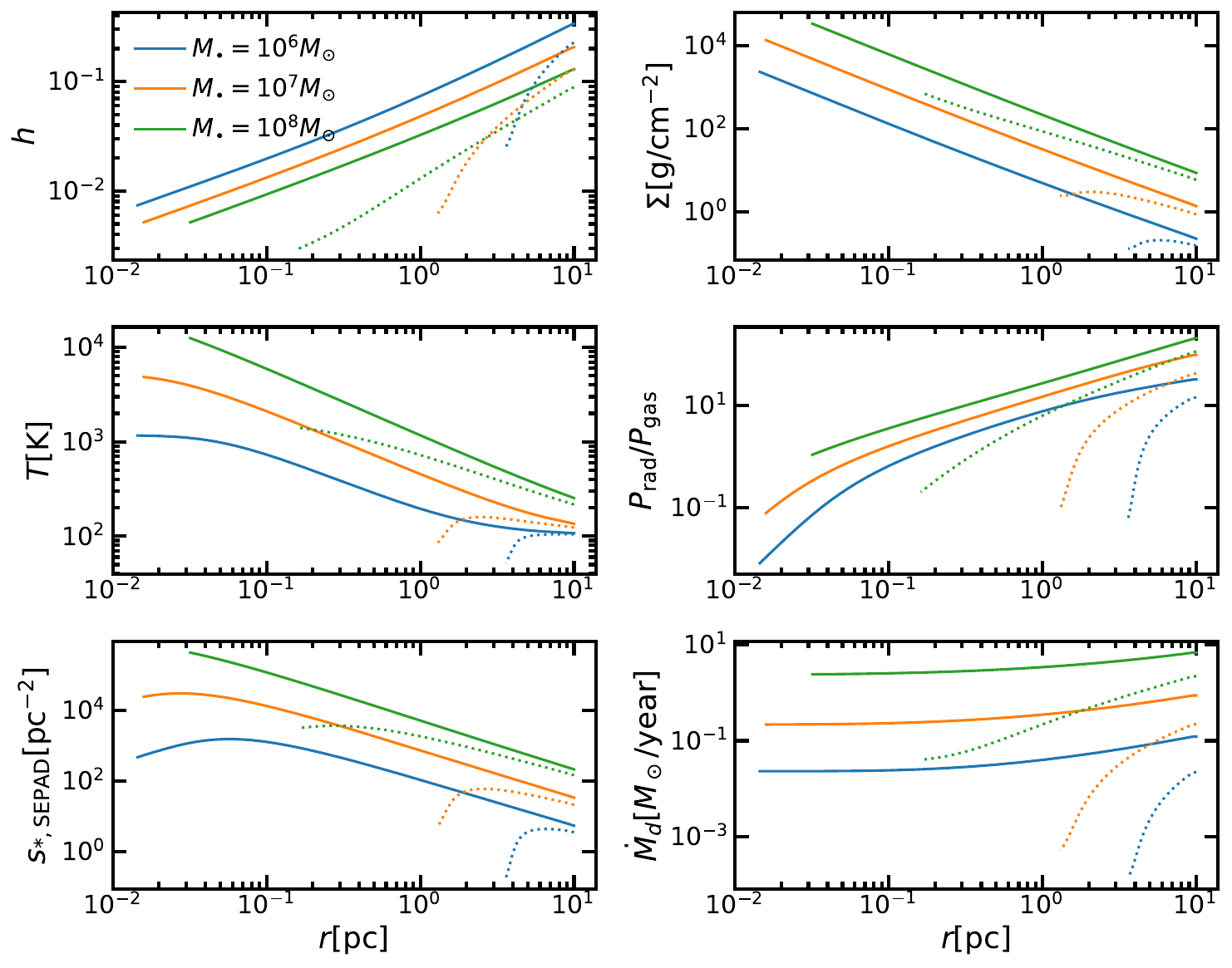}
\caption{The aspect ratio $h$, 
gas surface density $\Sigma$, temperature $T$, radiation to gas pressure $P_{\rm rad}/P_{\rm gas}$, equilibrium stellar surface number density $s_{*,\rm SEPAD} = Q^-/L_\star$, 
as well as the variable disk accretion rate as functions of radius $r_{\rm pc}$ for stars on the SEPAD track. 
We assume the radiative cooling is balanced by combined luminosity from massive stars with characteristic mass $M_*$ and luminosity $L_*$. 
We assume $\kappa_{\rm cgs} =  \alpha_\nu = 1$. 
The solid lines correspond to models with inner boundary accretion rate matched to $\tilde{\epsilon}_{\rm d} = 0.1$, 
while dotted lines correspond to outer boundary accretion rate matched to $\tilde{\epsilon}_{\rm d} = 0.1$.}
\label{fig:SBD_profile}
\end{figure*}

The decrease in mass accretion leads to an asymptotic profile that's reduced from 
Eqn \ref{eqn:s_*,IMS} by a factor of $\sim (1-f_{\rm deplete})^{\gamma}$. 
However, it is worth noting that due to helium enrichment in the stellar core
along the SEPAD track, 
their time-averaged $L_\star$ is lowered by a factor of $f_{L_\star}$ compared to immortal models which would increase $s_{*,\rm SEPAD}$ accordingly. 
Figure \ref{fig:SBD_profile} shows example numerical SEPAD disk models 
for different SMBH masses as well as two groups of accretion rates, adopting an average $L_\star = 2.2\times 10^6 L_\odot$ informed by stellar evolution models from \citet{alidib2023}, 
corresponding to $f_{L_\star} \sim 4$. 
For disk profiles corresponding to the dotted lines,
We still fix $\tilde{\epsilon}_{\rm d} (R_{\rm out})= 0.1$ 
as in Figure \ref{fig:SID_profile}. 
We assume $M_{\rm rem}\sim 30M_\odot$ \citep{alidib2023}
so $\epsilon_{\rm recycle} = 0.01$.
The recycle efficiency would be larger and 
$f_{\rm deplete}$
smaller with a lower residue black hole mass at the end of massive star evolution
$M_{\rm rem}$ (Eqn \ref{eqn:epsilon}).
Consistent with Eqn \ref{eqn:f_factor}, 
$f_d$ decreases with SMBH mass for a given $\epsilon_{\star}$ 
and $\tilde{\epsilon}_{\rm d}$.
For the $M_{\bullet} = 10^8 M_\odot$ model,
$f_{\rm deplete}\sim 1$ 
so a small fraction of $\dot{M}_{\rm d}(R_{\rm out})$ can flow into the inner parsec.
For lower SMBH mass models,
only the outer region is left with non-negligible surface 
density to contribute to star formation $\dot{\Sigma}_*$ 
and radiative cooling term $Q^-$, 
as well as small amount 
of AGN stars. The disk is markedly truncated 
within $R_{\rm trunc}$,
which satisfies 
$1-\sqrt{R_{\rm trunc}/R_{\rm out}} \sim 1/f_{\rm deplete}$, 
when $f_{\rm deplete}$ 
is larger than unity. 

Nevertheless, 
there is no reason for 
$\tilde{\epsilon}_{\rm d}$ 
at the outer boundary 
to be subject to rigid constraints since accretion feedback occur in the inner region of AGN disks. 
With somewhat larger ${\dot M}_{\rm d} (R_{\rm out})$, 
$f_{\rm deplete}$ would be markedly reduced and a much larger accretion rate fraction would reach the inner regions 
of the disk with ${\dot M}_{\rm d} (R_{\rm in}) \sim {\dot M}_\bullet$.
Solid lines in 
Figure \ref{fig:SBD_profile} correspond to disk profiles that iteratively select $M_{\rm d} (R_{\rm out})$ parameters, 
such that the inner region asymptotic $\dot{M}_{\rm d}$ is consistent with $\tilde{\epsilon}_{\rm d}(R \ll R_{\rm out}) = 0.1$. 
For $m_8=1$, 
$\dot{M}_{\rm d} (R_{\rm out})$ needs to be at least raised above the critical value by a factor of 3 
such that $f_{\rm deplete} < 1$. 
For $m_8=0.1, 0.01$ the rise in  $\dot{M}_{\rm d} (R_{\rm out})$ need to be more significant. 
Due to the increase 
in gas supply, 
these disks have gas and massive star density even higher 
than the corresponding IMS models in the outer disk, 
while the profiles converge at small radii save for a factor of $f_{L_\star}$. 

\subsection{Quantitative Velocity Dispersions and Merger Timescales}
\label{sec:merger_timescales}

\begin{figure}[htbp]
\centering
\includegraphics[width=0.42\textwidth,clip=true]{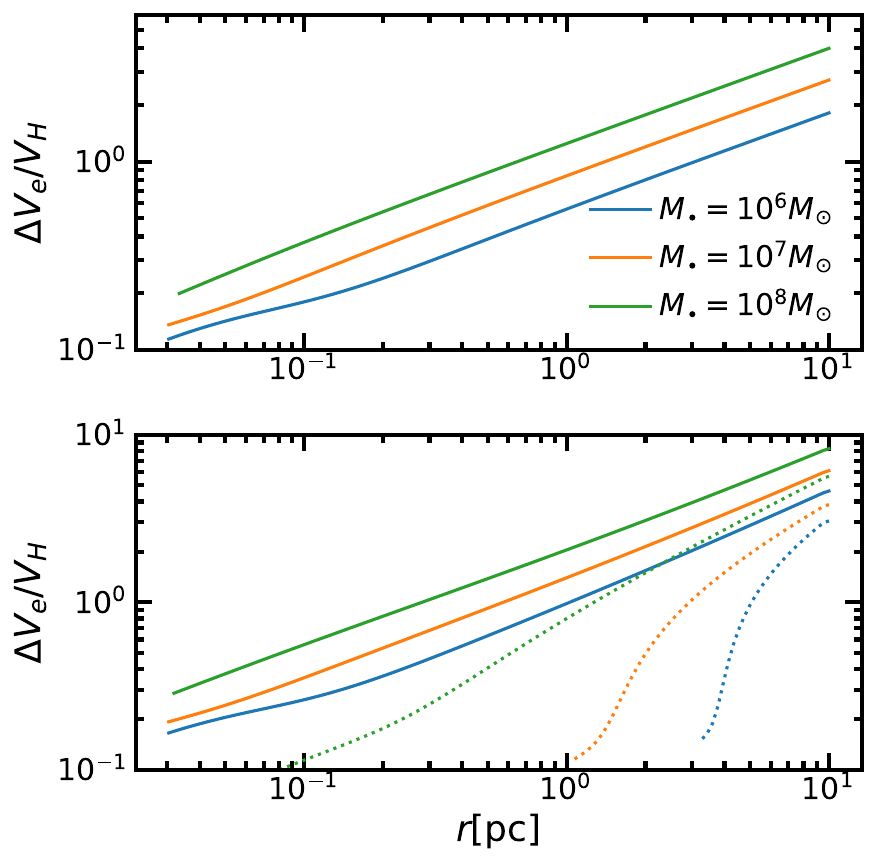}
\caption{$\Delta V_{e}/V_H$ on the IMS track for disk profiles 
presented in \S \ref{sec:ims_disk} (top panel) and the SEPAD
track \S \ref{sec:SEPAD_disk} (lower panel). When $\Delta V_{e}/V_H < 1$, the effective capture radius in Eqn \ref{eqn:final_capturetimescale} reaches its physical upper limit at $R_{\rm H}$. }
\label{fig:SID_veldis}
\end{figure}

The equilibrium number density of stars 
greatly influence their velocity dispersion and capture timescales. 
Informed by discussion in \S \ref{sec:capture_timescale}, 
we plot $\Delta V_e/V_H$ in 
Figure \ref{fig:SID_veldis} for the IMS (left panel, disk profiles corresponding to Figure \ref{fig:SID_profile}) as well as the SEPAD (right panel, 
disk profiles corresponding to Figure \ref{fig:SBD_profile}) disk models. 
{
In all cases, 
the effect of gas eccentricity damping is important 
in reducing the velocity dispersion amplitude from the escape velocity 
on the stellar surface (in the gas-free scenario) 
towards values close to the Hill velocity, 
especially so in the limit of very low $s_{*}$, 
corresponding to 
the inner regions of low accretion rate SEPAD models. When $\Delta V_e/V_H < 1$, the effective capture radius reaches its physical upper limit at $R_{\rm H}$.}

\begin{figure}[htbp]
\centering
\includegraphics[width=0.42\textwidth,clip=true]{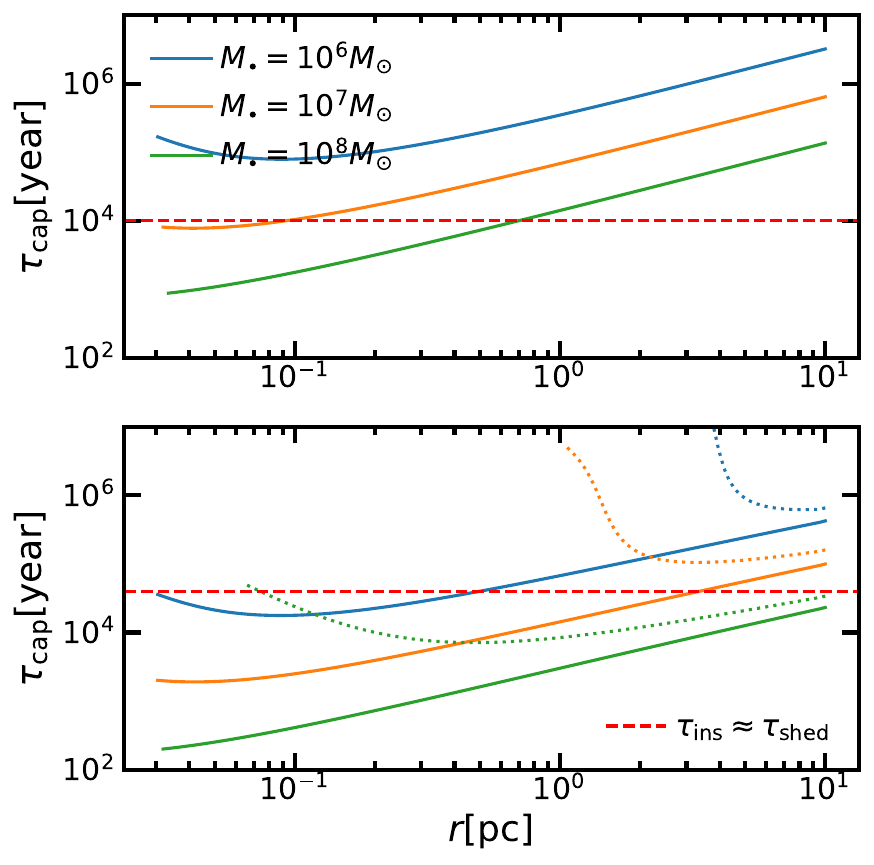}
\caption{Timescales of dynamical processes on the IMS track for disk profiles presented in \S \ref{sec:ims_disk} (top panel, disk profiles corresponding to 
Figure \ref{fig:SID_profile}) and on the SEPAD track in 
\S \ref{sec:SEPAD_disk} (bottom panel, disk profiles corresponding to Figure \ref{fig:SBD_profile}). 
The capture timescales $\tau_{\rm cap}$ are plotted in solid lines and dependent on $s_{\star}$ while the magnitude of $\tau_{\rm ins}\sim \tau_{\rm shed}$ is plotted in horizontal red dashed lines. 
The typical merger timescale should be approximately $\tau_{\rm cap} + \tau_{\rm ins}$. Note that $ \tau_{\rm ins}$ is longer by $f_{L_\star}$ in the SEPAD model due to the lower stellar luminosity.}
\label{fig:SID_timescale}
\end{figure}

In Figure \ref{fig:SID_timescale} we plot the capture timescales (solid lines) calculated from the forementioned velocity dispersions, 
against the estimate of the inspiral/shedding timescales (horizontal red dashed lines). 
In the IMS models (top panel), 
since stellar number density is high, 
the capture timescale $\tau_{\rm cap} \propto 1/s_{\star}$ is short, 
and for $M_\bullet \gtrsim 10^7 M_\odot$, $\tau_{\rm ins}$ dominates 
the merger timescale $\tau_{\rm merge} = \tau_{\rm cap} + \tau_{\rm ins}$ 
for certain regions where $s_*$ is large. 
In this limit,
hierarchical clusters, with a few members, may form. 
Since $\tau_{\rm ins}
\sim \tau_{\rm shed}$, mergers products can usually efficiently shed mass in between mergers and such that the characteristic stellar mass is marginally stable against runaway growth. 
Nevertheless, dispersion in $\tau_{\rm ins}$ and dynamical
evolution in the hierarchical clusters may still result in a fraction 
of stars going through runaway mergers and eventually acquiring sufficient 
mass to undergo collapse. 
The byproducts of collapse are either black holes 
with minimal gas return ($\epsilon_{\rm recycle} \ll 1$) or pair instability
supernovae with nearly all mass returning to the disk ($\epsilon_{\rm recycle} 
=1$). 

In the SEPAD models (right panel), 
if $\tilde{\epsilon}_{\rm d}(R_{\rm out})=0.1$ is constrained (dotted lines), 
the capture timescale becomes much longer due to the lower equilibrium $s_{\star}$ 
and therefore $\tau_{\rm cap}$ would always dominate the merger timescale, 
being the limiting timestep ($\tau_{\rm cap} \gg \tau_{\rm shed}, \tau_{\rm inspiral} $). 
In this case, 
the typical stellar mass is firmly stable against runaway growth. 
On the other hand, 
if the outer boundary accretion rate can freely increase to match $\tilde{\epsilon}_{\rm d}=0.1$ at the inner boundaries (solid lines), 
the radial dependence for $\tau_{\rm cap}$
would be similar to the IMS model. {A lower average luminosity results in a longer shedding timescale as well as larger stellar number density, 
so the inspiral timescale again becomes the limiting timestep in a merger event.}

\subsection{Mass Budgets and Potential Effect of Opacity Window}
\label{sec:mass_budget}

\begin{figure}[htbp]
\centering
\includegraphics[width=0.42\textwidth,clip=true]{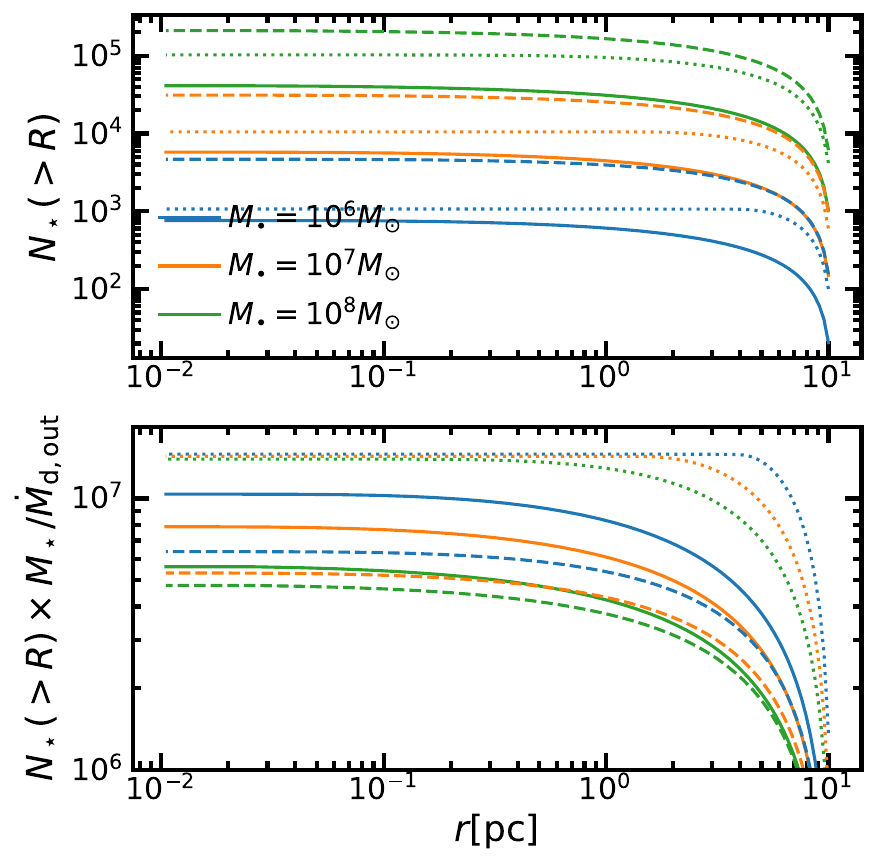}
\caption{Upper panel: the integrated number of stars beyond radius $R$ as functions of $R$ or $r = R/$pc, 
for the IMS (solid lines) as well as SEPAD models (dashed lines and dotted lines); Lower panel: the integrated number multiplied by the factor $M_\star/\dot{M}_{\rm d}(R_{\rm out})$,  which approximates the timescale 
towards establishing an equilibrium solution.}
\label{fig:number}
\end{figure}

{In Figure \ref{fig:number} 
we plot the integrated stellar number $N_\star(>R)$ 
for both the IMS (solid lines) and SEPAD (dashed and dotted lines) models, 
the asymptotic value of which at the inner boundary can reach of order $\sim 10^5 m_8$.
For such a high number density of AGN stars, 
it is also important to examine 
the mass budget. 
The lower panel shows $N_\star(>R) M_\star/ \dot{M}_{\rm d}(R_{\rm out})$,
providing an estimate of the timescale required for the accretion rate 
to be converted into this quantity of stars and achieve a quasi-steady state.
For the IMS models (solid lines), 
the formation timescale is of order $\lesssim 10^7$ years, 
a significant fraction of AGN lifetime ($\sim 10^8$ years),
suggesting a steady state can only be achieved at mature stages of AGN evolution.
The formation timescale 
is reduced for SEPAD models (dashed lines) where the outer accretion rate can increase to match $\tilde{\epsilon_{\rm d}}=0.1$ at the inner boundary, 
since the accretion rate at $R_{\rm out}$ 
is significantly 
enhanced while the stellar number density is increased by a smaller factor. 
In fact, 
it can be seen from Eqn \ref{eqn:s_*,IMS} that $s_\star$ only scales with $\tilde{\epsilon}_{\rm d}^{-1/3}$ 
while the mass accretion rate by definition scales with $\tilde{\epsilon}_{\rm d}^{-1}$,
such that a high accretion rate
shortens the formation timescale $\propto s_\star/\dot{M}_{\rm d} $ 
and mitigate the mass budget issue. 
In fact, 
if the accretion supply from the outer boundary diminishes over an AGN lifetime, 
the large $\tilde{\epsilon}_{\rm d}$ models may eventually transition to low $\tilde{\epsilon}_{\rm d}$ SEPAD models (dotted lines), 
as the disk gradually hollows out from the inside. 
With the formation timescale 
is further prolonged due to the reduced accretion rate, 
we may expect this depletion to escalate into a runaway process during late stages of AGN evolution. }

{Given that this formation timescale depends 
on the rate of mass supply, 
it is worth mentioning 
that realistic opacities can play a significant role in determining the upper limit of $\dot{M}_{\rm d}(R_{\rm out})$ for a fixed SMBH accretion rate.  
While we confirm that that 
applying a more realistic grain opacity treatment 
$\kappa \propto T^2$ below 100 K \citep{BellLin1994} 
only quantitatively affect our results, 
$s_{\star}$ becomes 
sensitive to the
opacity profiles around the sublimation temperature around
$T_{\rm sub} \sim 1000-2000$K. 
In this opacity window, 
if the midplane temperature $T_c$ is used to calculate $\kappa(\rho_c, T_c)$ 
and in turn the optical depth, 
radiative cooling becomes 
very efficient and the star formation rate must be exceptionally high to maintain energy balance. 
Therefore, 
even if $f_{\rm deplete} < 1$ (where $f_{\rm deplete}$ is determined by outer boundary opacities) 
and a large amount of accretion rate can flow into the inner disk, 
only a fixed amount of $\dot{M}_{\rm in}\sim 0.1-1  M_\odot $/yr, 
independent of $\dot{M}_{\rm d}(R_{\rm out})$, 
can penetrate further beyond the opacity window, and the rest 
would all be converted into an additional population of stars in a small annulus close to the sublimation radius \citep[][see their Appendix A]{TQM}. 
If such a steady state can be sustained, 
the stellar density in the outer disk could scale up with arbitrarily large $\dot{M}_{\rm d}(R_{\rm out})$, 
while the timescale towards achieving steady state in the outer disk 
would be shortened due to the aforementioned reason.}

{However, due to the extreme sensitivity of opacity to temperature in this regime, 
the vertical structure of the disk plays a crucial role in determining the vertically integrated optical depth, and the average opacity that controls cooling 
may significantly deviate from $\kappa(\rho_c, T_c)$. 
For example, 
the disk photosphere can host an opacity significantly larger than the midplane while having non-negligible density.  Such a thermal stratification would lead
to convective instability and an adiabatic vertical structure \citep{lin1980}.
Moreover, 
the temperature profile in the vertical direction
may be heavily influenced by irradiation \citep{chiang1997, garaud2007}
and accretion of stars when $s_\star$ becomes dense enough to cover the midplane. 
%The filling factor of stars $\sim s_\star R_{\rm H}^2$ increases significantly not only because of a large $s_\star$, but also because the disk is geometrically thin in this region such that $R_{\rm H}$ of $\sim 100M_\odot$ objects exceeds the disk scale height \citep{Dempsey2022}. 
While the details of this feedback process is subject to investigation through radiation hydrodynamic simulation, 
we chose not to quantitatively model the opacity window in this paper given this major uncertainty. 
We only note that
the overall stellar number density is already quite large in our estimates without the effect of opacity gap. 
Should the opacity window mechanism prove effective, 
our numbers for $N_\star$ 
would only be a conservative estimate for moderate values of $\dot{M}_{\rm d}(R_{\rm out})$ and can further increase for larger outer disk accretion rates.
}

\subsection{Integrated Merger Rates}

Taking $\tau_{\rm merge} = \tau_{\rm cap} + \tau_{\rm ins}$, 
the integrated merger rate beyond a certain radius $\Gamma_\star(>R)$ is given to be

\begin{equation}
   \Gamma_\star(>R) = \int_R^{R_{\rm out}} \dfrac{s_\star}{\tau_{\rm merge}} 2\pi R' dR' 
   \label{eqn:integrated_rates}
\end{equation}

\begin{figure}[htbp]
\centering
\includegraphics[width=0.42\textwidth,clip=true]{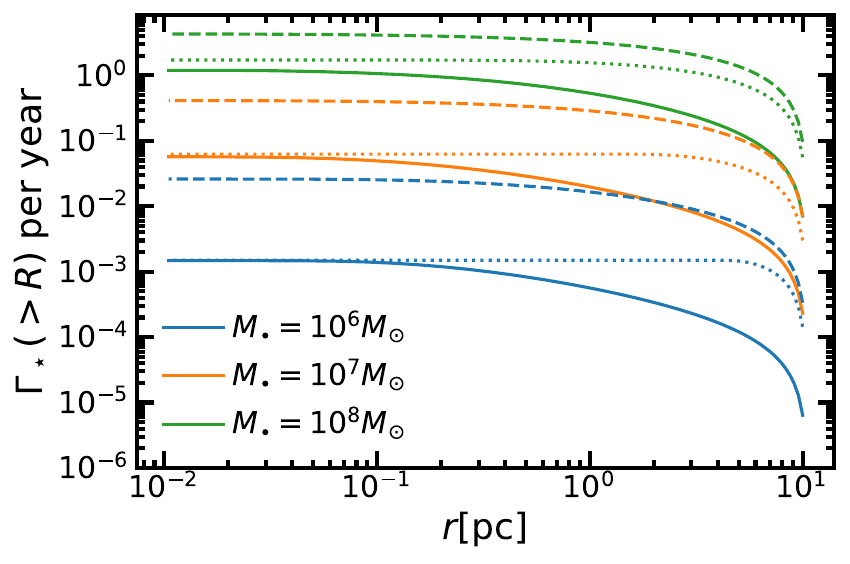}
\caption{The integrated/cumulative merger rates beyond radius $R$ as functions of $R$ or $r = R/$pc (Eqn \ref{eqn:integrated_rates}), 
for the IMS (solid lines) as well as SEPAD models (dashed lines and dotted lines).
}
\label{fig:rate_merge}
\end{figure}

In Figure \ref{fig:rate_merge} 
we plot $\Gamma_\star(>R)$ for both the IMS (solid lines) and SEPAD (dashed and dotted lines) models. 
The asymptotic values for $\Gamma_\star(>R)$ at small radii indicate that the choice of stellar evolution models can have large impact on the distribution of merger rates. 
For the IMS models, 
$\Gamma_\star(>R)$ strictly increases with SMBH mass for given $\alpha_\nu$ and $ \tilde{\epsilon}_{\rm d}$, with differential contribution from all disk radii. For the SEPAD models, 
if $\tilde{\epsilon}_{\rm d}=0.1$ 
is constrained at the outer boundaries (dotted lines), all contributions to $\Gamma_\star(>R)$ would be in the outer disk while the asymptotic values are similar to the IMS model. 
If $\tilde{\epsilon}_{\rm d} = 0.1$ is constrained at the inner boundaries (dashed lines), 
$\Gamma_\star(>R)$ can be notably larger than the IMS models due to larger $s_\star$ 
at regions close to $R_{\rm out}$.

%E.g. the overall merger rate $\Gamma_\star(>0)\sim 10^{-4}$ per year at $m_8 = 0.01$ or relatively low $f_{\rm deplete} \ll 1$, for both the IMS and the SEPAD model, while at $m_8 = 1$ and $f_{\rm deplete} \gg 1$, $\Gamma_\star(>0)$ varies from $10^{-3}$ to $3\times 10^{-5}$ per year for the IMS and the SEPAD model respectively.

\section{Summary and Discussions}
\label{sec:conclusion}

In this paper, 
we have derived equilibrium surface density profiles of massive stars in AGNs under two limiting contexts.
In the first scenario the stars are immortal (IMS track) 
during AGN's active phase and efficiently recycle back all accreted 
gas if it's not converted into luminosity, 
such that the disk exhibits nearly smooth disk accretion-rate profiles.
Alternatively, 
in the case where stars evolve off the main sequence into remnants (SEPAD track), 
their demise effectively removes significant amount of gas from the disk accretion flow and self-limit their number density. 
We define a physically motivated $\epsilon_{\rm recycle}$ factor to differentiate the effective removal of disk gas by stellar formation and evolution in the IMS and SEPAD scenarios (Eqn \ref{eqn:epsilon}), 
which also serves as a natural bridge between \citet{SirkoGoodman2003}-like 
and \citet{TQM}-like disk models. 
{This framework 
can be incorporated 
into state-of-the-art 1D disk modeling tools such as 
those constructed by \citet{Gangardt2024}}.

The observed populations of 
young stars near the Galactic center where 
(mostly B-type) 
S-stars with semi-major axes 
$a_\star \sim 5\times 10^{-3}-0.04$pc
\citep{genzel2003, Ghez2003a, gillessen2017} are 
surrounded by a population of clockwise-rotating disk 
(mostly O/WR type) stars with $a_\star \sim 0.04-0.5$ pc,  
\citep{levin2003, lu2006, yelda2014, vonfellenberg2022}.
A zone of avoidance highlights the absence of stars
with pericenter distance and eccentricity smaller than
those of the S stars \citep{burkert2024}.  
This observed
feature is more consistent with
the possibility that these young stars formed coevally in 
SEPAD-type disk models, 
with ${\dot M}_{\rm d}$ increasing with $R$ (Zheng et al. in prep). 
Additionally, 
we comment that a ``hollow" disk due to star formation depletion can also possibly lead to low-luminosity AGNs such as NGC 3167 %, once a candidate of ``true'' type II AGNs 
\citep{bianchi2019}.

Focusing on constraining number density and merger rate of stars, 
we apply a simple constant-opacity disk model that captures the essential features of the stellar-luminosity-heated outer region of AGNs. 
Integrating the stellar surface density, 
we estimate that the total number of AGN stars within a few parsecs to be $N_\star \sim \mathcal{O}(10^3) -  \mathcal{O}(10^5)$.
These objects can encounter each other and merge on a timescale of $\tau_{\rm merge}$, 
which ranges from $\mathcal{O}(10^4) -  \mathcal{O}(10^6)$ years, 
while the lower limit is set by the shortest timescale for binaries to harden under the effect of surrounding gas. 
{We also estimate that for $M_\bullet$ between $10^6-10^8M_\odot$, 
the overall merger rates per AGN ranges from $\Gamma_\star \sim 10^{-3}$ to $\sim 1$ 
per year.  This inference places some constraints in the 
rate of massive star mergers in large-scale surveys with $\sim 10^6$ AGN samples such as ZTF \citep{Frederick2021} \& LSST \citep{LSST2019} to be $\sim \mathcal{O}(10^3-10^6)$ per year. Typical timescale is determined by details of the 
energy dissipation of the collision process \citep{hu2024accretion}, which is currently poorly understood. 
It may range from as short as the dynamical timescale 
to as long as the shedding timescale.} 
Our calculation provides motivation for detailed hydrodynamic studies of massive mergers in AGN disks, 
which could help find traits to identify them from the plethora of transients. 
When mutual inclination 
between a massive star and the accretion disk is excited
during the inspiral, 
the star can also collide and shock-heat the disk during each vertical passage and produce Quasi-periodic Eruption(QPE)-like signals \citep{Linial2023,Tagawa2023}.

An important characteristic of SEPAD stars is their contribution to the enrichment of the AGN environment. 
As they transition off the main sequence and evolve into stellar mass black holes (sBHs), 
each SEPAD star redistributes approximately $\Delta M_{Z} \sim$ a few solar masses of $\alpha$ elements back into the disk. 
Consequently, 
the overall pollution rate can be estimated as ${\dot M}_{Z} \sim \pi R^2 \Delta M_{ Z} s_{\rm \star, SEPAD}/ \tau_\star$. 
This production rate of heavy elements results in a significant increase in metallicity, 
with $\Delta Z \sim {\dot M}_{Z} /{\dot M}_{\rm d}$, which is several times higher than the solar value \citep{alidib2023} and consistent with observational data \citep{Huang2023}. On a timescale $\tau_{\rm AGN} \sim 
10^8$ year, the cumulative surface density of produced sBHs grows to
$s_{\rm BH} \sim s_{\rm \star, SEPAD} \tau_{\rm AGN}/\tau_\star$ and they eventually becomes the dominating heating source.
The Eddington-limited rate of accretion onto these embedded 
black holes leads to ${\dot \Sigma}_{\rm BH} \ll {\dot \Sigma}_{\rm rem}$
in Eqn (\ref{eq:sigmadiffusion}), significantly reducing the stellar contribution needed to maintain 
a thermal equilibrium at late stages of AGNs, shutting off star formation (Eqn \ref{eq:energyequi}). In contrast, IMS stars neither contribute to the metallicity enrichment nor evolve into sBHs.

It has been suggested that merging black holes detected by LIGO Virgo 
collaboration may find one another in AGN disks \citep{McKernan2012,McKernan2014,Tagawa2020a,Samsing2022}, 
with their merging process possibly associated with electromagnetic (EM) counterparts \citep{graham2020,Tagawab2023}. 
Since disk capture of compact objects from the nuclear star cluster is inefficient due to the small cross section \citep{Wang2018,wang2022}, the
SEPAD track provides a promising channel for the formation of these gravitational wave sources. \citet{ChenX2020} suggested that if sBH mergers occur inside a gaseous medium, 
the gravitational wave signal can be altered by gas dynamics, 
which provides another way of directly identifying merger in gas-rich channels apart from EM counterparts.
Additionally, 
if sBHs and massive stars can coexist in large numbers within an AGN disk,
massive stars may efficiently capture sBHs successively on a similar timescale as the stars' own $\tau_{\rm cap}$ (since the impact parameter is dominated by the star's Hill radius), 
which then merge inside the core of the massive star after common-envelope evolution. 
In this scenario, 
gas dynamics would distort the
GW signal even more significantly due high density in the stellar core.
This kind of merger scenario is first proposed to explain possible electromagnetic counterpart to event GW150914 \citep{Loeb2016,Dai2017}, 
and expected to be rare for field stars due to their short lifetimes. 
However, such events can be prevalent in AGN disks. 

In subsequent works on detailed time-dependent AGN disk models, 
details of stellar evolution should be taken into account. 
It is conceivable that the disk may undergo regulated cycles between the two limiting solutions, 
{if the stellar number/heating distribution is allowed to advect radially under effect of migration \citep{Grishin2023, Wu2024}. 
The capture of stars from the surrounding nuclear cluster can also be added as a source term for the stars at small radii \citep{Wang2023,Generozov2023}. 
In SEPAD disk model where 
stars evolve into sBHs, 
the effect of accretion from sBHs should be taken into account in the heating and density depletion terms \citep{Gilbaum2022,Tagawafeedback2022,ChenKen2023,zhou2024stellar}.}
Moreover, 
our calculation of binary capture and inspiral are subject to numerous uncertainties. 
To resolve these uncertainties and provide more accurate estimation of stellar merger rates, 
detailed 
investigation of migration and capture processes in a dynamically crowded disk 
using hydrodynamic and/or N-body simulations is warranted. 

\begin{acknowledgements}
%\vspace{5mm}
We thank Xian Chen, Adam Dempsey, Jamie Law-Smith, Jiaru Li, 
Hui Li, Shengtai Li, Man Hoi Lee 
and Jeremy Goodman for helpful discussions.

\end{acknowledgements}

\bibliography{sample631}{}
\bibliographystyle{aasjournal}



\end{document}